\PassOptionsToPackage{table}{xcolor}
\documentclass[acmlarge]{acmart}

\usepackage{cite}
\usepackage{algorithmic}
\usepackage{graphicx}
\usepackage{textcomp}
\usepackage{xcolor}
\usepackage{lipsum}
\usepackage{hyperref}
\usepackage{multirow}
\usepackage{comment}
\usepackage{cleveref}
\usepackage{svg}
\usepackage{tablefootnote}
\usepackage{tabularx}
\usepackage{enumitem}

\setcopyright{rightsretained}
\acmJournal{DTRAP}
\acmYear{2024} \acmVolume{1} \acmNumber{1} \acmArticle{1} \acmMonth{1}\acmDOI{10.1145/3688806}

\begin{document}
	
\title{On NVD Users' Attitudes, Experiences, Hopes and Hurdles}
\subtitle{Authors' version; to appear in ACM DTRAP Special Issue on IMF 2024}
	
\author{Julia Wunder}
\affiliation{
    \institution{IT Security Infrastructures Lab, Friedrich-Alexander Universität Erlangen-Nürnberg (FAU)}
    \city{Erlangen}
    \country{Germany}
}
\email{julia.wunder@fau.de}

\author{Alan Corona}
\affiliation{
    \institution{IT Security Infrastructures Lab, Friedrich-Alexander Universität Erlangen-Nürnberg (FAU)}
    \city{Erlangen}
    \country{Germany}
}
\email{alan.corona@fau.de}

\author{Andreas Hammer}
\affiliation{
    \institution{IT Security Infrastructures Lab, Friedrich-Alexander Universität Erlangen-Nürnberg (FAU)}
    \city{Erlangen}
    \country{Germany}
}
\email{andreas.hammer@fau.de}

\author{Zinaida Benenson}
\affiliation{
    \institution{IT Security Infrastructures Lab, Friedrich-Alexander Universität Erlangen-Nürnberg (FAU)}
    \city{Erlangen}
    \country{Germany}
}
\email{zinaida.benenson@fau.de}

\renewcommand{\shortauthors}{Wunder et al.}

\begin{abstract}
The National Vulnerability Database (NVD) is a major vulnerability database that is free to use for everyone. It provides information about vulnerabilities and further useful resources such as linked advisories and patches.
The NVD is often considered as the central source for vulnerability information and as a help to improve the resource-intensive process of vulnerability management. Although the NVD receives much public attention, little is known about its usage in vulnerability management, users’ attitudes towards it and whether they encounter any problems during usage.
We explored these questions using a preliminary interview study with seven people, and a follow-up survey with 71 participants. The results show that the NVD is consulted regularly and often aids decision making. Generally, users are positive about the NVD and perceive it as a helpful, clearly structured tool. But users also faced issues: missing or incorrect entries, incomplete descriptions or incomprehensible CVSS ratings. In order to identify the problems origins, we discussed the results with two senior NVD members. Many of the problems can be attributed to higher-level problems such as the CVE List or limited resources. Nevertheless, the NVD is working on improving existing problems.
\end{abstract}



\maketitle

\section{Introduction}
\label{sec:introduction}

The number of vulnerabilities in the form of created Common Vulnerabilities and Exposures (CVEs\footnote{\url{https://www.cve.org}})
has been continually rising in the last years and there are no signs of a slow down \citep{statista-cves}. The complexity in managing this enormous amount of vulnerabilities, constantly checking for relevance and planning mitigations, rises along with the number of CVEs. The National Vulnerability Database (NVD\footnote{\url{https://nvd.nist.gov}}) 
plays a central role in vulnerability management and information distribution and is commonly considered the standard 
for vulnerability information. For this purpose, users expect extensive, well-documented and most importantly correct information. Lesser quality in this regard may lead to misinformation and in turn wrong decisions, prioritizing less severe or less relevant vulnerabilities over more severe or more relevant ones and thus may cause damage and weaken overall security. To avoid this, the database does not only need to be correct, but also usable and understandable.

As a vulnerability knowledge base and aggregator, the NVD is the central point of information regarding vulnerabilties' severity, details and affected systems. 
It facilitates decisions and resource planning concerning priority.
However, little is known about usage of this central service. Related work is commonly focused on vulnerabilities themselves or on strategies for improving vulnerability management  
(see \Cref{sec:related-work} for examples and details), while concrete insight into users' experience and daily work interacting with the NVD is yet to be researched.

\paragraph{Contribution} The contributions of this paper are as follows:
\begin{itemize}
    \item We provide insight on how the NVD is used in everyday work, e.g., it is consulted on a near daily basis to get an overview about specific vulnerabilities, used as a reference to prioritise issues and to find patches.
    \item We show that the NVD is perceived mostly positive by its users, as many praise its standardized and well-structured look. However some problems exist, such as incorrect or incomplete NVD records, and the users need to find ways to circumvent them, e.g., by verifying or completing the information through usage of other sources.
    \item
    We shed light on the potential origins of the problems that users face when consulting the NVD by discussing them with two senior members of the NVD. For example, incomplete NVD records can be attributed to problems further upstream in the CVE Program.
\end{itemize}

\paragraph{Outline} The paper is structured as follows. First, we provide background for the NVD, discuss related work and derive research questions in \Cref{sec:background}. We then describe the preliminary interview study in \Cref{sec:prelim-study}. Subsequently, in \Cref{sec:main-study} we provide the design and the results of the main study. In \Cref{sec:nvd-interviews} we present the insights gained from interviewing two senior employees of the NVD. We discuss the findings in \Cref{sec:discussion} and conclude in \Cref{sec:conclusion}.
\section{Background and Research Questions}
\label{sec:background}

\subsection{The National Vulnerability Database}

\subsubsection{The CVE Program}
To understand the NVD, one has to know about the CVE Program, which is a
standardized method to identify and classify publicly known vulnerabilities, assigning each vulnerability a unique identifier called a CVE ID \citep{cve-faq}.
Vulnerabilities are collected and catalogued in the publicly accessible CVE List as CVE Records since 1999 by the MITRE\footnote{\url{https://www.mitre.org}} corporation \citep{cve-history}.
MITRE is a non-profit organization partially funded by the US \citep{mitre-history}, whereas the CVE Program is funded solely by the US.
CVE IDs and thus CVE Records are assigned by MITRE and CVE Numbering Authorities (CNAs), a group of researchers and vendors from different countries, who can reserve CVE IDs \citep{nvd-and-cve}. 

After a vulnerability has been discovered and reported to a CNA or MITRE, it is then verified by those organizations and a new CVE ID is requested or an already reserved ID is assigned. The CVE Record is enriched with details, submitted and then published in the CVE List \citep{cve-lifecycle}.

A CVE Record must contain the following elements \citep{cve-requirements}:
\begin{itemize}
    \item The corresponding CVE ID, which contains the year the vulnerability was publicly disclosed or reserved by a CNA and a continuous number for each year, e.g., CVE-2022-35260 for a vulnerability from the year 2022 \citep{cve-id}.
    \item A description summarizing the vulnerability, which must contain the affected product name with version number and vulnerability type, root cause or impact \citep{cve-requirements}. Further information may be added such as the product's vendor, 
    the attack vector, the type of attacker or state of remediation \citep{nvd-and-cve}.
    \item References containing complementary information such as links to papers, press articles, patches or similar, that deal with the vulnerability.
\end{itemize}
\subsubsection{From the CVE Program to the NVD}
\label{sec:background-nvd}
The National Vulnerability Database is a collection of information about security vulnerabilities and was created in 2005 by the National Institute of Standards and Technology (NIST\footnote{\url{https://www.nist.gov}}) \citep{nvd-history}.
NIST is part of the U.S. Department of Commerce \citep{nist-about}, thus both the NVD and the CVE Program are funded by the United States.

NVD employees analyze each CVE Record after it has been published in the CVE List.  Every record is expanded with information and published in the NVD, while the content of the original CVE Record is retained \citep{nvd-and-cve}. The API offered by the NVD enables users to query relevant entries automatically.

A NVD record contains the following information \citep{nvd-record-details, nvd-and-cve}:
\begin{itemize}
    \item The CVE ID, description and references of the corresponding CVE List Record. The references are enhanced with manual Internet research and categorized with labels such as ``Patches'', ``Issue Tracking'' and others, whereas the ID and description remain unchanged.
    \item The severity of the vulnerability according to the Common Vulnerability Scoring System (CVSS\footnote{\url{https://www.first.org/cvss}}). The resulting score is a number between 0.0 and 10.0 and is intended to make severities of vulnerabilities comparable with each other. It is important to note that a CVSS score is not supposed to be used as a risk score, but rather as additional information that must be applied in the local system to appraise the local risk of a vulnerability \citep{cvss}.
    \item If the vulnerability is in the Known Exploited Vulnerabilities Catalog (KEV Catalog\footnote{\url{https://www.cisa.gov/known-exploited-vulnerabilities-catalog}}) of the Cybersecurity and Infrastructure Security Agency (CISA\footnote{\url{https://www.cisa.gov}}), this fact will be displayed. The KEV Catalog contains vulnerabilities that are actively exploited according to CISA \citep{kev-catalog}.
    \item The Common Weakness Enumeration (CWE\footnote{\url{https://cwe.mitre.org/about/index.html}}) names that apply to the vulnerability, which represent different categories, such as ``CWE-20: Improper Input Validation'' \citep{cwe}.
    \item The information about which software or combinations of software are affected, written in the Common Platform Enumeration (CPE\footnote{\url{https://nvd.nist.gov/products/cpe}}) naming scheme. This scheme classifies products according to various product categories, operating systems and hardware units \citep{cpe}, making it possible to search the NVD for these identifiers. 
\end{itemize}

In summary, the NVD takes each published vulnerability entry of the CVE List and adds information before publishing it, while retaining the original information of the CVE Record. The NVD takes on the role of an aggregator that structures the information, enriches it with details and distributes it.

\subsection{Related Work}
\label{sec:related-work}
The related papers are categorized into works that use the NVD as a basis for their research (\Cref{sec:related-work-databases}), for example investigating the information flow of vulnerabilities, and papers that have researched the NVD for possible flaws and improvements (\Cref{sec:related-work-critics}).

\subsubsection{Vulnerability Management and Databases}
\label{sec:related-work-databases}

Vulnerability databases such as the NVD provide essential information on vulnerabilities for security experts. Thus, it is crucial to understand how the vulnerability information finds its way to the analysts.
\citet{2020-miranda} analyzed the information flow of vulnerabilities and the role of various vulnerability databases. They found that although the NVD is not necessarily the first database to publish the vulnerabilities, in most cases it is the second one to present them.
According to the authors, the NVD  clearly takes the position of an aggregator. Miranda et al. highlight the relevance of the NVD, as it is one of the largest platforms contributing to the dissemination of vulnerability information.

Despite being considered as a significant database in literature, the NVD does not seem to list all vulnerabilities, as \citet{2022-forain} noted when comparing the NVD with other databases, such as the China National Vulnerability Database\footnote{\url{https://www.cnvd.org.cn}}.
\citet{2020-nerwich} also observed that IoT-specific vulnerabilities are often not taken into account in vulnerability databases such as the NVD.

\citet{2019-zhang} researched the correlations between vulnerabilities and features of affected systems, for example the system complexity, to predict future vulnerabilities using machine learning and based their analysis on data of the NVD.
\citet{2018-williams} developed a model to investigate how vulnerabilities change over time so that potential gaps in the NVD and other vulnerability databases could be identified and trends predicted.
\citet{2011-zhang} also examined whether future release dates of vulnerabilities in certain systems can be predicted using existing information of the NVD. However, the results showed that the NVD records are a rather weak source for this approach, as the entries are often of poor quality according to the authors.

\subsubsection{NVD Critics and Improvement}
\label{sec:related-work-critics}

Besides related works that use the NVD as a basis for their research, there were also works that explicitly examined the NVD for possible flaws.

\citet{2013-nguyen} analyzed whether versions of Google Chrome listed as having vulnerabilities in the NVD were really vulnerable. To do this, they examined 539 vulnerabilities for 12 Chrome versions and extracted the commit logs and determined whether the vulnerability has been fixed. The authors found that 25\% of the versions examined were incorrectly specified in the NVD.
Similar problems were also noted by \citet{2019-dong} who developed an automated system to detect inconsistencies between CVE descriptions and  NVD records. In particular, the authors highlighted that the version numbers given for vulnerabilities are often incorrect. For example, software versions are sometimes included in the NVD record that are not actually affected by the vulnerability. Even worse, in the case of some particularly critical vulnerabilities, software versions that are affected by the vulnerability are completely missing. 
\citet{2022-anwar} also systematically analyzed NVD records and presented a system for automatic correction, as many entries are prone to errors. In particular, entries for vendor names are at risk of errors, as variations of names are often used for the same vendor. Anwar et al. therefore introduce a vendor heuristic to tackle this problem.

In addition to the incorrect software versions, there were other problems, as \citet{2018-rodriguez} discovered. The authors investigated vulnerability publication delays and compared publication dates of the NVD with other vulnerability databases such as ExploitDB\footnote{\url{https://www.exploit-db.com}}. The authors found that the majority of vulnerabilities are published with a delay of one to seven days in the NVD, although there were also vulnerabilities that had a delay of more than 200 days. Vulnerabilities such as buffer errors, cross-site scripting and information leak disclosures had a particularly long delay. However, the authors also point out that the NVD publishes a new vulnerability as soon as it has been included in the CVE List and that the NVD subsequently adds further information within two days.

\citet{2022-guo} also noted that problems with delays are not necessarily due to the NVD itself, but to the CVE database upstream. In their work, they first examined vulnerabilities in the CVE List and then compared these with NVD records. They found that many incomplete entries in the CVE List were completed in the NVD. To support this manual process, they then developed a tool that uses machine learning to predict missing entries, such as vulnerability type or attack type, with a high probability.
An expanded analysis of the NVD using machine learning and natural language processing techniques was also conducted by \citet{2021-kuehn}. The authors developed an automated system that can evaluate the information quality of vulnerability descriptions in the NVD. The information quality was measured by three factors: accuracy, completeness and uniqueness. The system was trained with a sample of 3,000 vulnerabilities that have been labeled by two IT security experts using tags, e.g., using the tag ``AV'' when the phrase ``attack vector'' occured. Based on the description provided by the NVD, the system can generate a CVSS vector string. These newly generated vector strings are then compared with the CVSS evaluations of the NVD to evaluate the accuracy of vulnerability descriptions.
The results showed that many vulnerabilities in the NVD do not match the expected vector string calculated automatically.

Similar to these related works, we also examine the NVD for possible flaws and inconsistencies -- but in contrast we focus on a human factors side. We shed light on how the NVD is used in daily work, what potential problems its users encounter and what improvements they might wish for. We also examine where these potential problems could arise from by discussing the findings with two senior NVD members and talk about possible improvements.
\subsection{Research Questions}
\label{sec:research-questions}
While the NVD is often stated as a prominent source 
when working with security advisories and vulnerabilities (\Cref{sec:related-work-databases}), little is known about its usage in everyday work of security analysts and other professionals. 
Furthermore, although related work revealed that some entries within the NVD are incorrect or incomplete (\Cref{sec:related-work-critics}), the problems that users encounter have not been researched so far. We aim to validate the insights of related work and to look further for problems and general usability of the database.
By asking users about their general attitude, we aim to gain insight on aspects previously unknown.
Finally, while every system has room for improvement, we investigate what aspects are most important to improve, either due to lack of quality or functionality, or because users would benefit most in a particular area. 
\begin{enumerate}[itemindent=2em,labelsep=0.5em,label=RQ\arabic*:]
    \item How is the NVD used in everyday work?
    \item What problems do users encounter when using the NVD?
    \item What is the attitude of users towards the NVD?
    \item Which improvements would be most valuable to users?
\end{enumerate}

Research questions RQ1-RQ4 are explored in the qualitative preliminary interview study and the quantitative main survey study.
To investigate the reasons behind and possible solutions to the issues encountered in RQ2, 
we discussed our findings with two senior NVD members to include the view of the provider.
\section{Preliminary Study - Interviews}
\label{sec:prelim-study}

\subsection{Study Design}
\label{sec:prelim-study-study-design}
To gain deeper insight into the use of the NVD and problems users might face, interviews were conducted in a preliminary study. The interview guide was divided into three sections, the first being demographic questions retrieving information such as age or job title of the interviewees.
Next, we asked about the usage of the NVD in daily work and concluded the interview by asking about problems the interviewees might have experienced. Three test interviews were carried out to improve the initial interview questions.
The interviews were conducted in Germany, in the native language, other languages or countries were out of scope of the study.
Consequently, all citations in \Cref{sec:prelim-study-results} were translated from German to English.

The study was approved by the data protection office of the Friedrich-Alexander-Universität
Erlangen-Nürnberg (FAU). Before beginning the interview, we provided the participants with an informed consent form that notified them about the study's purpose and their rights of information and deletion. Participation was voluntary, and participants could interrupt or cancel at any time without any disadvantages. Participants could also inform us if certain information should not appear in the transcript. There was no compensation. The answers were pseudonymized. The recordings were stored on a server of our university, the data was only accessible internally to authorized persons.

The interviews were conducted remote via Zoom
and recorded with OBS\footnote{\url{https://obsproject.com}}.
Afterwards, the recordings were fully transcribed, in the beginning manually with oTranscribe\footnote{\url{https://otranscribe.com}} and later automatically with Whisper\footnote{\url{https://github.com/openai/whisper}}. Transcripts were edited, removing sensitive statements, for example if interviewees said the name of their employer or named internal tools,
and inductively coded with MaxQDA\footnote{\url{https://www.maxqda.com}} by one coder. In the process of coding, the codebook was discussed in several meetings with two additional team members and adjusted accordingly. The codes were then aggregated by category and the interviews analyzed across cases.

\subsection{Participants}
\label{subsec:prelim-participants}
The target group of the study were people who regularly use the NVD.
In order to find participants, we used personal contacts and social media such as X and Reddit. LinkedIn was also utilized to search for organizations and people that are active in the IT security sector.
Additionally, all members of the German CERT Alliance\footnote{\url{https://www.cert-verbund.de}} were contacted.
In total, seven people from Germany were interviewed, the demographics can be found in \Cref{table:prelim-demographics}. 
The interviewees were between 24 and 54 years old, all except one were male. All interviewees except one have a university degree. 
The job titles of the interviewees vary greatly, but all of them work in IT security. P2 and P7 work for the same company A, and P3 and P5 work for the same company B, which means that the interviewees come from a total of five different organizations with 50 to 10,000 employees.

{\rowcolors{1}{lightgray!30}{white}
\begin{table}[!htbp]
    \centering
    \begin{tabular}{c m{0.4cm} m{1cm} m{1.6cm} m{2.9cm} m{1.4cm} m{2.8cm} m{2.2cm} }
        \bottomrule
     ID & Age & Gender & Education & Job title & \mbox{IT work} \mbox{experience} & Specialization & Employees\\
        \toprule
       P1 & 34 & Male & PhD & Referent & 16 years & Threat Intelligence & 100 -- 300\\
       P2 &  54 & Male & Diploma & Senior Security Consultant & 20 years & Security Management & 50 - 100 (A)\\
    P3 & 24 & Male & IT Specialist & Security Engineer & 5 years & Application Security & \mbox{1,000 - 10,000 (B)}\\
      P4 &  38 & Male & PhD & Professor, Information Security Governance & 15 years & \mbox{Teaching and} \hspace{0.9cm} Research & 1,000 - 10,000\\
      P5 &  33 & Diverse & PhD & Security Engineer & 9 years & Product Development & \mbox{1,000 - 10,000 (B)}\\
      P6 &  46 & Male & M. Sc. & Principal & 27 years & IT Security Advisor & 300 - 1,000\\
      P7 &  35 & Male & B. Sc. & IT Security Advisor & 4 years & Security Management & 50 - 100 (A)\\
    \end{tabular}
    \caption{Demographics of the interviewees. P2 and P7 work for the same company A, P3 and P5 work for the same company B.}
    \label{table:prelim-demographics}
\end{table}
}
\subsection{Results}
\label{sec:prelim-study-results}

\subsubsection{NVD Usage}
All interviewees use databases to gather information. Five of them consult databases of the German CERT Alliance, databases of manufacturers, and Linux distributions are consulted as well. Open Source and manufacturer programs are used for static and dynamic code analysis, as well as scanning included libraries, to discover possible vulnerabilities.
Some of these programs obtain all the information from the NVD.
Social media, such as X, are being scanned automatically with scripts, albeit P6 mentioned that the IT Security community on X is slowly disintegrating. Print media, mailing lists, and personal exchange were mentioned as well by P2 and P6.

Four interviewees use the NVD regularly to enrich their security advisories. In addition, interviewees consult the NVD to research details about vulnerabilities, such as taking the CVSS score as a reference or validating information found in other sources. Only one respondent uses the API of the NVD directly, while others employ the API integrated into tools.

\subsubsection{Issues when Using the NVD}
All participants except P1 have experienced issues when consulting the NVD.
Four interviewees often do not agree with the assigned CVSS scores, as they seem to be too high or not matching the vulnerability.
Errors in NVD records were encountered by P2, P4 and P7 such as wrong CPE 
identifiers and swapped or ambiguous descriptions. CVE Records not being validated before the publication in the NVD is perceived as a problem as well.
The period of time until a NVD record is analyzed by the NVD staff was criticized as too long by four interviewees. P2 even reported a waiting time of over a year. 

Three interviewees criticized the NVD's CPE naming scheme, for example that it is confusing how the dependencies between software products are propagated in their CPE identifiers. 
According to P4, CPE identifiers cannot work because they consist of version numbers and some companies do not assign version numbers to their products. Another criticized aspect was that the NVD takes too much time to adjust CPE identifiers when one company is acquired by another.

The problem of varying quality of NVD records was also mentioned as an issue, because CNAs 
are only audited once. P4 summarized this problem:

\begin{quote}
    \emph{``The quality [of NVD records] varies greatly depending on the team that fills it out.
    So there are teams that have incredibly good security experts, people with
    20 years of experience [\dots] and there are teams that write in the minimum that can be written, or
  the information is often wrong because it hasn't been
    validated.'' }
\end{quote}
P4 and P5 see the search functionality of the NVD as cumbersome when trying to search for specific entries or CPE identifiers. 
Further problems were identified: manufacturers sometimes disagreeing with contents of NVD records, missing details such as references, and CVSS, according to the participants, being a poor standard due to its subjective nature. 

Due to the problems with the NVD, four interviewees obtain or validate information about vulnerabilities additionally from other sources.
The company of P2 takes the issues with the CVSS scoring of the NVD as an opportunity to carry out its own vulnerability scoring, for example by examining the source code of the vulnerability.

\subsubsection{Trust in the NVD}

Four interviewees expressed their trust in the NVD because they have had good experiences with the database so far. P2 and P3 trust  the NVD since it is widely used by many people.
Furthermore, P1 and P3 justify their trust by trusting NIST or the US government. P2 stated:

\begin{quote}
    \emph{``[The] NVD is the database par excellence for vulnerabilities. 
    Everyone uses it and that wouldn't be the case if the trust [\dots] wasn't there.''}
\end{quote}

Two interviewees mistrust the NVD since they do not trust any source unconditionally. P6 mistrusts the NVD because of the influence of the USA, P4
because of its problems and P1 due to vulnerabilities not being validated.

\subsubsection{Perceived Positive and Negative Aspects of the NVD}
\paragraph{Positive aspects.}
The interviewees appreciate the existence of the NVD, especially the CVE ID and CVSS scores. For example, P2 said:

\begin{quote}
    \emph{``In any case, the positive thing is that there is a broad overall view of vulnerabilities. Someone has to do it so that one knows what one's talking about and that you have the information readily available. So that's definitely really, really great.''}
\end{quote}
The quality of the NVD content is praised by three interviewees, its open nature and references by P4 and P6. The API, stability of the CVSS scoring and standardized format of data are also perceived as positive.
P4 reminisced:

\begin{quote}
    \emph{``Back then, vulnerabilities were exchanged by fax or as e-mails with poor text files. And now you have a public database where everything is transparent and where there is a process behind it that is accepted by everyone. So I think this is an extremely positive achievement that we have.''}
\end{quote}

\paragraph{Negative aspects.}
Two interviewees criticize the NVD's national and incomplete nature. For P4, the automation of the NVD API is insufficient, as it only returns information or messages about vulnerabilities, but does not apply patches. 
There is a desire for fully automated system, as it can significantly reduce the required resources.
Another negative aspect is that the NVD is a single point of failure if the used tools obtain their information directly from the NVD.

\subsubsection{Sympathy with the NVD}
Some interviewees expressed a positive attitude towards the NVD and offered constructive criticism.
Four interviewees put their previous negative statements into context. For example, it was noted twice that mistakes happen because of human nature or that problems occur due to the enormous amount and complexity of vulnerabilities.
According to P2 and P5, incomprehensible CVSS scores happen because the NVD has to issue a rating for each system without knowing every specific application case.
P2 and P4 showed understanding with the problems, as the NVD does not have enough employees in their eyes.
According to P4, the NVD is taken for granted and not appreciated enough. In the same sentence, the NVD is called upon to develop further. 
P4 summarized:

\begin{quote}
    \emph{``But I can tell you one thing, what is being done behind the scenes, what money is needed for [the NVD], how many people are working on it, also working on it voluntarily.
    That is immense.''}
\end{quote}

\subsubsection{Takeaways for Main Study}
The preliminary study showed that the NVD is being used regularly and is well-received by its users. The trust of interviewees in the NVD is high and several positive aspects were noted. However, negative aspects were also mentioned. Further investigation is required to determine the extent of these problems and to identify any additional issues that may need attention. The preliminary study helped to set up the questionnaire for the main study, for example to define questions about the exact use of the NVD and the reasons behind it. It also enabled to identify initial problems that could be quantitatively measured in the survey.
\section{Main Study - Survey}
\label{sec:main-study}

\subsection{Study Design}
\subsubsection{Questionnaire and Testing}
\label{sec:questionnaire}

Based on the results of the interviews, the online survey was created to examine the research questions quantitatively. The questionnaire was divided into five themes. First, the participants were asked whether they use the NVD in everyday life. Participants who answered ``No'' to this question were provided additional questions about whether they were generally familiar with the NVD and what other sources they use to obtain vulnerability information. The survey was then closed for these participants, as we are focusing on people who use the NVD regularly.

Next, participants who regularly use the NVD 
were queried about their usage of the NVD, for example
for what reasons they use the database. There were also questions about which aspects of an NVD record are being used and how important they considered them to be. We then asked the participants to briefly describe in a text field how the information in the NVD is further processed in their everyday work.
The next part of the survey was concerned with potential problems that users may encounter when using the NVD and the frequency of these problems, if they have already experienced them. 
In the third part, we asked about the participants' personal attitude towards the NVD and provided
statements about usability and attitudes, such as ``The NVD is easy to use'' or ``I feel dissatisfied with the NVD''. Participants were able to indicate their agreement with these statements using a 5-point Likert scale from ``strongly agree'' to ``strongly disagree''.
The participants also had the opportunity to explain in a text field what they like and dislike about the NVD. This was followed by a question about possible improvements. Finally, demographic questions were asked.

The tool LimeSurvey was used to implement the online survey. A test phase with five people was conducted and their feedback incorporated. The testers were either from the IT security research sector or have been working in IT security for many years.

\subsubsection{Ethics}
\label{sec:ethics}

The study has received approval from the data protection office of the Friedrich-Alexander-Universität
Erlangen-Nürnberg (FAU). At the beginning of the survey, an informed consent was presented to the participants, that notified them about the purpose of the study and their rights of information and deletion. Only those who consented were able to take part in the survey. Participation was voluntary and participants were able to interrupt or cancel the survey at any time without any disadvantages. There was no compensation. The answers were pseudonymized, the participants are referred to in following as PS$<$number$>$. The survey was hosted on a server on-site, and only authorized personnel had access to the data.

\subsubsection{Data Analysis}
\label{sec:data-analysis}

The data was extracted as JSON files from LimeSurvey and analyzed descriptively with purpose-built Python scripts.
Considering the free text fields in the survey, we decided to use an intercoder agreement based on the recommendation by \citet{2019-mcdonald}, as there was a possibility of inaccurate interpretation.
Two team members coded the free text answers independently and created codebooks. These codebooks were discussed and merged into one, which was then used again to code the answers. We used Cohen's Kappa $\kappa$ to see how well the coding matches \citep{1977-cohen}. Overall, the calculated agreement was over 0.7 to 0.9 for most of the codes, which indicates a substantial to almost perfect agreement \citep{1977-landis}. All differences were thereafter discussed to reach full agreement.

\subsubsection{Recruitment}
\label{sec:recruitment}

We advertised the survey on thematically appropriate Subreddits\footnote{For example \url{https://www.reddit.com/r/cybersecurity}}, mailing lists such as Fulldisclosure\footnote{\url{https://nmap.org/mailman/listinfo/fulldisclosure}} and used personal contacts. We also asked each participant to share the survey with other relevant people.
In total, 187 participants started the survey. Some participants did not agree to the informed consent form or did not finish the survey, which results in 101 
valid responses. These participants can be categorized into two groups -- those who use the NVD in their daily work (N = 71) and those who do not (N = 30). Since the target group consists of people who consult the NVD regularly, the demographic data and results are based on their responses (N = 71). Participants who do not use the NVD in their daily work were asked if they were generally familiar with it and what other sources they use (see \Cref{sec:results-usage} for more information). On average, the survey took 15 minutes to complete.

\subsection{Participants and Demographics}
\label{sec:participants}

{\rowcolors{1}{lightgray!30}{white}
\begin{table}
	\begin{center}
		\begin{tabularx}{0.8\linewidth}{l|X X}
			\toprule
			& $N$ & $\%$ \\
			Total & 71 & 100.0 \\
            \bottomrule
			Male & 52 & 73.2\\
			Female & 2 & 2.8\\
			Diverse & 2 & 2.8\\
			N/A & 15 & 21.1\\\hline
			18-29 years & 4 & 5.6\\
            30-39 years & 16 & 22.5\\
			40-49 years & 21 & 29.6\\
			Above 50 years & 11 & 15.5\\
			N/A & 19 & 26.8\\\hline
			Germany & 34 & 47.9\\
			Other European country & 10 & 14.1\\
			USA & 11 & 15.5\\
			Other country & 3 & 4.2\\
			N/A & 13 & 18.3\\\hline
			No academic education & 5 & 7.0\\
			Bachelor's degree & 12 & 16.9\\
			Master's degree & 31 & 43.7\\
			Ph.D. & 11 & 15.5\\
			Other & 2 & 2.8\\
			N/A & 10 & 14.1\\\hline
			1-5 years work experience & 17 & 23.9\\
			6-10 years work experience & 19 & 26.8\\
			11-20 years work experience & 21 & 29.6\\
			More than 20 years work experience & 9 & 12.7
		\end{tabularx}
		\vspace*{0.5em}
		\caption{Overview of the participants' demographics and work experience of the main study.}
		\label{tab:demographics}
	\end{center}
\end{table}
}

\Cref{tab:demographics} shows an overview of the demographics. 
The vast majority of the participants are male (73\%) and are on average 41 years old. Approximately 48\% of the participants come from Germany, 16\% from the USA and 18\% from other countries, although 18\% did not provide any information here. Most of the participants have a Master's degree (44\%), some have a Bachelor's degree (17\%) or Ph.D. (16\%).

Considering their work environment, 73\% of the participants are regular employees, and 11\% 
are self-employed. On average, they have been working in IT security for around 12 years. Almost all of them indicated a job title that can be assigned to the IT security field, such as security consultant, security analyst or penetration tester.

About 32\% of the participants work in the information and communication sector, while 11\% work in professional, scientific or technical fields and about 7\% in education. The remaining participants are spread across sectors such as financial areas, manufacturing, transportation and others.
About 21\% work in a company with more than 5,000 employees, 20\% with 1,000-4,999, 14\% with 250-999, 16\% with 50-249 and 17\% with 1-49 employees. This indicates that companies from different industry sectors and of different sizes consult the NVD regularly.
\subsection{Results}
\label{sec:results}

\subsubsection{NVD Usage}
\label{sec:results-usage}

The NVD is used regularly by 71 out of 101 participants. 
Overall, 93 participants stated that they were familiar with the NVD, which underlines the popularity of the database. People who do not use the NVD on a regular basis (N = 30) referred to security advisories, social media, mailing lists, search engines or other databases such as VulnDB\footnote{\url{https://flashpoint.io/ignite/vulnerability-management/}} as their source for vulnerabilities. 
The following results are based on the answers of the 71 participants who use the NVD in their everyday work.

\begin{figure}
	\begin{center}
            \includegraphics[width=0.75\textwidth]{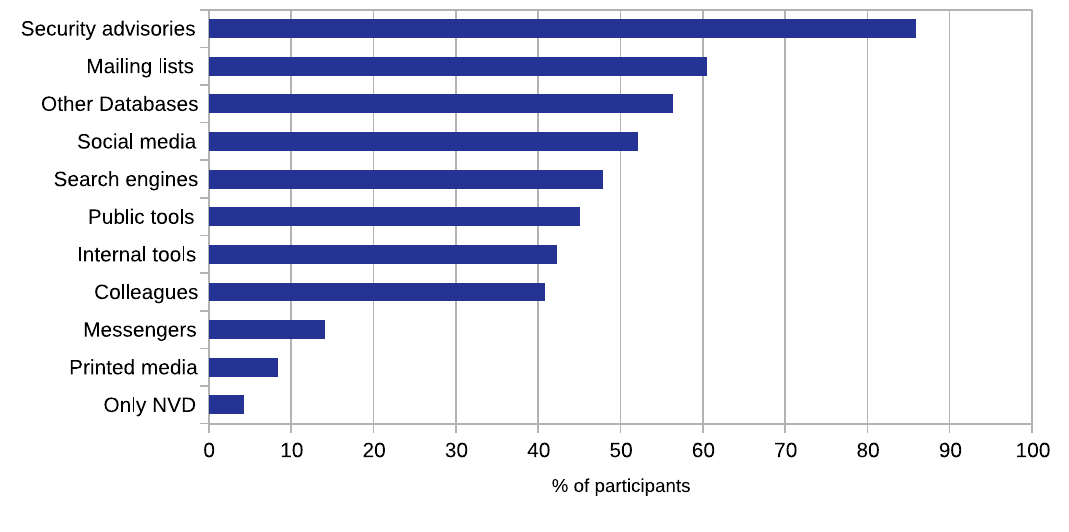}
	\end{center}
	\caption{Sources that are used by the participants for gathering information about vulnerabilities ($N = 71$).}
	\label{fig:NVDsources}
\end{figure}

On average, participants have been consulting the NVD for 6.7 years (median 5). Most use it on a weekly basis (42\%) or even daily (39\%). Although the NVD has been accessed with high frequency for several years, it is not the only source used by participants (see \Cref{fig:NVDsources}). Only about 4\% rely solely on the NVD for information on vulnerabilities. Most participants also use security advisories (86\%), mailing lists (61\%) or other databases (56\%), similar to the participants who do not regularly use the NVD. Social media, such as X, is also used for finding vulnerability information (52\%), though this has decreased over the past year. 
35\% of these participants stated that with the re-branding from Twitter to X, they no longer use the platform. 

Despite the fact that the vulnerability information is collected from different sources, most of the participants (80\%) have one main source from which they obtain the information. For 47\% of participants, 
the NVD is the primary source. This shows that the NVD is quite relevant in the context of vulnerability management, as it is used by many 
as the main source to obtain vulnerability information.

The NVD is most often used to collect more details about a vulnerability (97\%) and to take
the CVSS score as a reference (70\%). It is also commonly utilized to validate the same vulnerabilities from a different source (61\%) or compare similar vulnerabilities (37\%). Other participants expand their own security advisories (37\%).

\begin{figure}
	\begin{center}
		\includegraphics[width=0.75\textwidth]{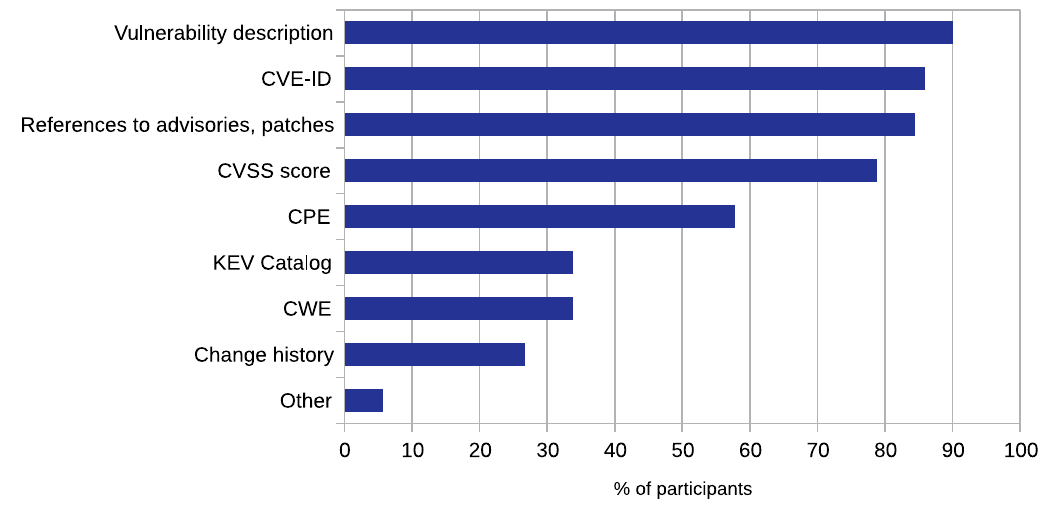}
	\end{center}
	\caption{Overview of which components of an NVD record are used by the participants ($N = 71$).}
	\label{fig:NVDentryUsage}
\end{figure}

An NVD record consists of various components (see \Cref{sec:background-nvd}). \Cref{fig:NVDentryUsage} provides an overview of components that are utilized. In particular, the vulnerability description (90\%), the CVE ID (86\%) and references to advisories and patches (85\%) are used most commonly. The CVSS score is also widely considered as a reference (79\%), aswell as the known affected software configurations (CPE) (58\%). We also asked the participants to rank the components of an NVD record in order of importance to them. For most participants, the CVE ID is the most relevant component, followed by the vulnerability description, the CVSS score and other references to security advisories. After the information has been collected, most of the participants (92\%) verify it by cross-checking it against other sources.

About 72\% of the participants used a free-text field to describe how the information from the NVD is subsequently processed in their company. It is most frequently used for filtering, for example to identify which of one's own systems are affected by the vulnerability. A decision is then made on how to prioritize the vulnerability based on the CVSS score. Linked patches and other information also supports the decision. The NVD is also widely used as a reference to provide a comparison for internal vulnerability assessments. Many participants describe the NVD as a well-structured overview and that is often used as a starting point: \emph{``[The] NVD is typically a jumping off point for investigation. We use it to get an overview of something we are about to deep dive on,''} PS18 stated.
Some participants explained that the information is incorporated into internal vulnerability assessment tools for further processing. Other participants prepare the collected information and forward it to affected customers. About 28\% of the free text answers explicitly stated to consult other sources to supplement or verify the information.
The participants justify this with the frequently missing or incorrect information: \emph{``NVD information is typically not detailed enough to determine the severity or exploitability of a vulnerability, so I supplement it with [other sources],''} PS54 explained.
Approximately 20\% of the participants stated that they adopt some information 
directly and without modification, such as the vulnerability descriptions.

\subsubsection{Issues when Using the NVD}
\label{sec:results-issues}

\begin{figure}
	\begin{center}
		\includegraphics[width=0.9\textwidth]{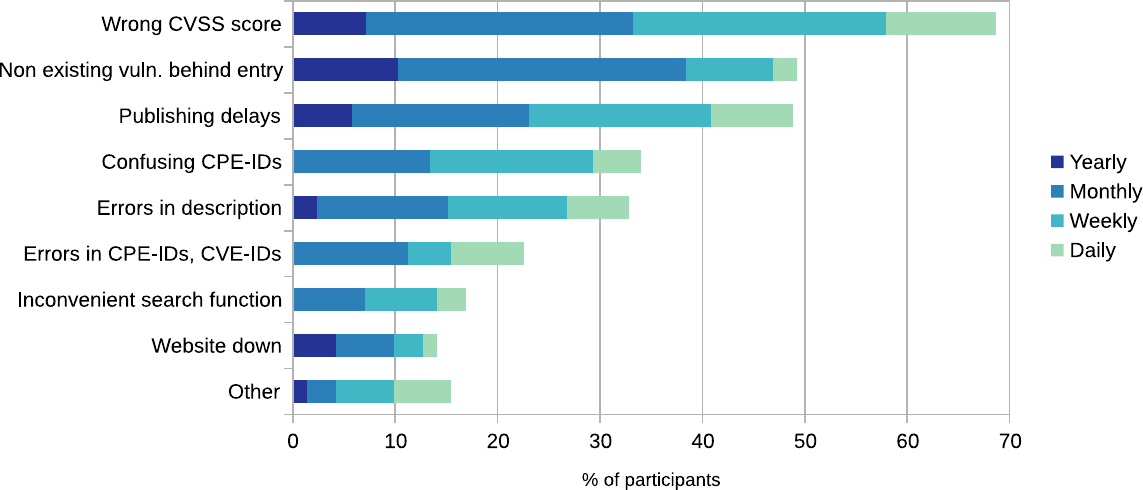}
	\end{center}
	\caption{Problems and their frequency participants encounter when using the NVD ($N = 71$).}
	\label{fig:NVDproblems}
\end{figure}

The results indicate that the NVD is consulted very frequently and in different ways by many people. However, only about 8\% of the participants stated to never have experienced any problems. \Cref{fig:NVDproblems} provides an overview of problems with the NVD and their frequency.

Approximately 68\% of the participants stated that the CVSS score does not match the vulnerability described in the NVD record and does not adequately reflect its severity in their eyes. Another problem according to 49\% of the participants is that there are very long delays in the publication of a new vulnerability and that NVD records are often published even though the vulnerability does not actually exist. Incorrect descriptions or confusing CPE IDs were also mentioned (34\%). Other problems were noted in the free text answers, such as criticism that the API is not reliable and too slow, as it frequently fails with error messages. The most frequent comment was that information in NVD records often contain errors, are outdated and incomplete, even though some vulnerabilities were published years ago. Sometimes the information is completely absent.
The mentioned problems usually occur weekly or monthly.
Overall, the data from the main study confirms the problems that we identified in the interviews from the preliminary study (\Cref{sec:prelim-study-results}).

\subsubsection{Attitude Towards the NVD}
\label{sec:results-attitude}

Although users regularly encounter problems, the NVD is still perceived as mostly positive.
In an optional free text field, 60\% of participants took the opportunity to describe what they personally see as good or bad about the NVD. 
Many comments praised the NVD for its good and clear structure, usability, and being free of charge. That the NVD is a standardized database which everyone uses as a guide was particularly emphasized positively. PS33 summarized this with: \emph{``It is awesome to have a central system where everything is. It would be a pain, if anybody releases there [sic] 
vulnerabilities in an unorganized way.''} However, some also criticized that the NVD is centralized and under the full authority of the US government.
PS1 therefore suggested to \emph{``have a crowdsourced system where we can all contribute data and get to better quality.''}

Many participants described the quality of NVD records as poor. As the previous results already showed, missing or incorrect CVSS scores, vulnerability descriptions or CVE IDs are highlighted in particular. Another negative aspect is the long delays and the lack of revisions: \emph{``It seems like nobody is checking if anybody really updates the information there,''} PS12 stated.
In addition, there is often a shortage of further details within the NVD records, which is why participants 
have to rely on other sources.
According to the participants, the CVSS score is perceived as particularly negative, due to it often reflecting a different level of severity than appropriate in their eyes.

\begin{figure}
	\begin{center}
		\includegraphics[width=0.9\textwidth]{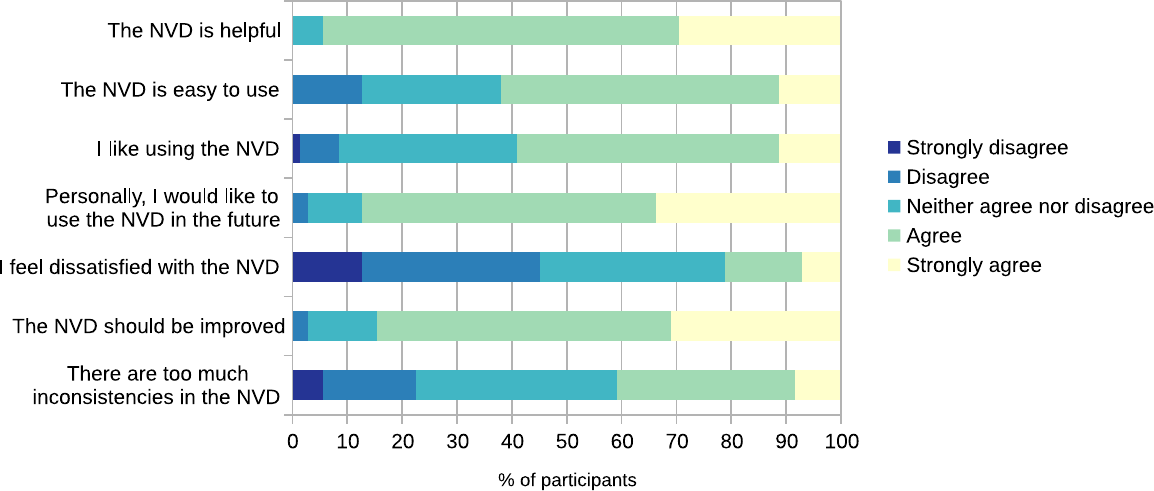}
	\end{center}
	\caption{Statements and participants' agreement for NVD usability ($N = 71$).}
	\label{fig:NVDusability}
\end{figure}

An overview of agreement with the statements about the NVD is given in \Cref{fig:NVDusability}. Most see the NVD as a helpful tool and find it easy to use. Many participants agree that they like using the NVD and will continue to do so in the future. However, the majority of participants stated that there are too many inconsistencies within the NVD and that it should be improved.

In terms of possible improvements, most participants  (56\%) would like the NVD to become international so that parties apart from the US government
can influence the database.
Many also think that the NVD records should be reviewed better (51\%) and that the existence of a vulnerability should be verified  before it is published (45\%).
Around 45\% of the participants would like to see faster publication times. In addition to the predefined response options, the participants could write further improvements as comments. Suggestions were API improvements, such as better integration into their own systems, and increased website stability.
It should also be more convenient to determine how a vulnerability affects a certain system.
Another room for improvement is the cooperation with CNAs. \emph{``Instead of reviewing entries before publication, [the] NVD must ensure that CNAs have high standards. If low quality CNAs grow, [the] NVD will never be able to keep up''}, PS39 stated.

Overall, similar to the preliminary study, users voiced their sympathy with the NVD. Many participants attenuated their criticism by acknowledging that humans are working at the NVD and human errors happen. Due to predetermined structures and a lack of external resources, it is difficult to solve these problems.
As PS48 summarized it: \emph{``NVD, CISA, MITRE and FIRST are great, but they can't do everything, and they all move slow.''}
\section{NVD Interviews}
\label{sec:nvd-interviews}

In order to identify the origins of the issues identified in the studies, we reached out to the NVD. 
In the following parts we present the NVD’s perspective on the problems we discovered.

\subsection{Study Design and Participants}

Two interviews were conducted, where the results were discussed with two senior NVD members: 
Tanya Brewer, who has been working as NVD Program Manager for four years, and Christopher Turner, who has been working as NVD Senior Advisor for nine years. Brewer provides strategic vision and top-level guidance regarding the NVD.
Turner is responsible for internal development decisions of the NVD, has efforts focused on external collaborations with vendors as well as CNAs and participates in various working groups. The first interview was conducted with Brewer and Turner simultaneously and a follow-up interview was conducted only with Turner.

The interviews were approved by the data protection office of the Friedrich-Alexander-Universität
Erlangen-Nürnberg (FAU). At the beginning, an informed consent was presented to the interviewees that notified them about the purpose of the study and their rights of information and deletion. Participation was voluntary and both interviewees were able to interrupt or cancel the interview at any given time without any disadvantages. The interviewees gave us their consent to state their names and job titles. There was no compensation. The recordings were stored on a server of our university, the data was only accessible internally to authorized persons. 
The interviews were conducted, recorded, transcribed, coded and analyzed in the same manner as the interviews in the preliminary study (\Cref{sec:prelim-study-study-design}).
The two interviewees were recruited via the personal network of an interviewee from \Cref{subsec:prelim-participants}.

The first interview was conducted after the preliminary study in November 2023 and was divided into three sections, starting with introductory questions that asked interviewees to talk about their professional role. Afterwards questions about problems of the NVD in the interviewees eyes and about the problems discovered in the preliminary study were asked.
This was followed by asking for causes and possible solutions, and concluded by supplementary questions about future plans that the interviewees can share with us. 

The second interview was conducted after the main study in April 2024.
As we did not discover a lot of new problems but rather confirmed the issues found in the preliminary study, the interview focused on what has changed since the last interview and whether the known problems have been addressed accordingly.
There have not been any significant changes in the NVD since the first
interview, consequently, the results in this section are based mainly on the
findings of the first interview.
After the second interview, the NVD requested to review the citations in the following section before publication. There were no objections. 

\subsection{Problems of the NVD}
\paragraph{Problems Stated by Interviewees.}
According to Brewer \emph{``there is human error, there is changes in philosophy [and] outright mistakes that haven't been caught''} as well as \emph{``a [\dots] problem with the quality of the data''}. This shows that the NVD is aware of the varying quality of NVD records. 
During the discussion, Turner highlighted the existence of information gaps in NVD records 
prior to 2015. Many policies, procedures, and practices have undergone significant changes around that time, which contributed to these gaps. He sees CPE maintenance as a point of concern for users because of consistency or data availability issues. Regarding historical issues, Turner noted that the CPE dictionary has been problematic due to its management by multiple organizations before it was transferred into the NVD's care. Brewer also mentioned that there have been slowdowns and outages since the organization began using a newer API.

Furthermore, Brewer described the working relationship with CISA as challenging at times
due to their frequent change of leadership.
On a side note, according to Brewer, the NVD Servers are hosted on AWS\footnote{\url{https://aws.amazon.com}} and internally managed partially by the NVD and partially by the NIST CIO office. 

\paragraph{Shortcomings of the CVE Program.}

Regarding errors in NVD records 
and quality differences, Brewer sees the CVE
Program responsible for that, as the NVD relies on the data provided by the CVE
List.
This data is sourced from various CNAs, leading to a variation in the
quality of the entries supplied.
\emph{``From one CNA to another, what gets entered can vary wildly'',} Brewer
stated.
According to Turner, the CVE Program guidelines for CNAs are too loose, and it is the CVE Program's responsibility to validate CVE Records. 
He stated that the NVD needs to focus on its own tasks to avoid getting
\emph{``spread indefinitely thin and unable to accomplish anything''}.
Furthermore, the failure of CNAs to disclose vulnerabilities in tandem with the
CVE Program leads to vulnerabilities being published before they are included in
the CVE List.
Despite these problems, Brewer described the cooperation with the CVE
Program as \emph{``fairly close''}.
Brewer summarized:
      \emph{``We have our role that we have carved out for ourselves and [\dots] pretty good reasons, why we stick to that role.''}
    
\paragraph{NVD Analysis Delays.} 
Brewer perceives the delay problem as not severe, as she believes that the NVD team has effectively managed to keep the number of not analyzed CVE Records 
consistently low, despite the growing number of emerging vulnerabilities.
Turner thinks that CNAs are responsible for delaying analysis at times because they publish vulnerabilities on an ad hoc basis.
This results in a high quantity of disclosed vulnerabilities waiting to be analyzed within a short period of time.
Additionally, it takes a long time to determine the CPEs for each NVD record. 

\paragraph{Shortage of Personnel and Resources.} 
Brewer mentioned that the staffing levels at the NVD have not been able to keep up with the steady increase in the number of disclosed vulnerabilities.
The reason for this is that the NVDs \emph{``budget has been flat or lessening over the last three years,''} as has every budget in the federal government. Additionally, Turner provided an insight into how exhausted NVD employees are:
\begin{quote}
     \emph{``[The NVD team is] at capacity for what humans can do within the work load and the expectations [\dots] and it is not even close to what is really necessary to be able to keep up with new publications, the needs of CPE management or updates to existing records after initial analysis.''}
 \end{quote}

\subsection{Possible Solutions and Future Plans}
Two possible solutions were suggested to address the issues with CNAs and the CVE List. 
Turner referred to CVMAP\footnote{\url{https://nvd.nist.gov/vuln/cvmap}}, a solution that is already in place.
CVMAP is similar to a light touch auditing and aims to encourage coordination, collaboration, and discussion between NVD analyst operations and CNAs.
As a second solution, Brewer mentioned the Vulntology\footnote{\url{https://pages.nist.gov/vulntology}} project developed by Turner, which is a framework for providing foundational vulnerability information in such a way that CNAs do not have to provide information they deem as confidential.

According to Brewer, there are existing plans to include more information in NVD records, 
such as Exploit Prediction Scoring System Scores\footnote{\url{https://www.first.org/epss}}, CVSS 4.0 scores\footnote{At the time of the interviews, the highest assigned CVSS version was v3.1}, and CPE data from external sources.
In addition, there are plans to update the CPE specification and to include persistent URLs (PURLs) in NVD records, 
which are web resource links that provide permanent access to a specific resource \citep{purl}.
This will enable users to map PURLs to Software Bills of Materials (SBOMs), which are machine-readable inventories that contain information about the software components and dependencies of a product, as well as their hierarchy and other details \citep{sbom}.
With this, NVD records could become easier to use and simpler to apply in one's own system
\section{Discussion}
\label{sec:discussion}

\subsection{Usage of the NVD (RQ1)}
The results of the studies show that the NVD is commonly consulted in everyday work and is often seen as a reference to fill own security advisories with information or make decisions. It is also frequently used as a starting point for obtaining further information on updates or linked advisories. Users praise the NVD's well-structured overview and the fact that it functions as a free, main database that everyone has agreed on and can refer to.
The results indicate that the NVD is quite relevant in the context of vulnerability management, as it is used by many as the main source to obtain vulnerability information. In this way, the NVD clearly fulfills its role as an aggregator of vulnerability information. 

\subsection{Issues and their Origin (RQ2)}
One aspect that is widely criticized by NVD users are CVSS scores that do not match the vulnerability description. However, this problem might be due to the subjective nature of CVSS itself \citep{2020-allodi,2018-spring,2024-wunder} 
and also due to the fact that the NVD analysts often do not receive enough details about the vulnerabilities from the vendors. 
Another problem are incorrect or incomplete NVD records, 
the quality of which has also been criticized in the past \citep{2011-zhang,2022-anwar}. The issue does not originate from the NVD, as 
in most cases, the NVD adopts the existing information from the CVE List directly and enhances it with additional details obtained through further research and analysis, which was also investigated in the past \citep{2022-guo}. 
This process often takes time, that consequently leads to delays of
vulnerabilities being published, which is criticized by users.
However the long time it takes to disclose a vulnerability is mainly caused by the CVE List.
According to related work \citep{2022-guo, 2018-rodriguez}, the NVD publishes a vulnerability usually within two days after it appears in the CVE List.

Overall, most of the issues arise from the adopted entries of the CVE List, which are of varying quality. 
Despite the fact that there
exist standards for the entries, they are not strict enough, which
leads to  differences in quality, depending on the vendor.
Thus, more precise standards are needed. 
There are approaches using machine learning to correct missing entries or incorrect information within the NVD \citep{2021-kuehn, 2022-anwar}. However, there is a risk that the models will be trained on poor quality vulnerability descriptions. 
This is the root of the problem and
could be solved by a standardized format or language for vulnerability
descriptions.
The NVD tries to accomplish
this task with their Vulntology project. 
Even though they would like to tackle these issues with full force, they are not
able to due to a shortage of staff and resources.

\subsection{Positive Attitude but Room for Improvement (RQ3, RQ4)}
Although users are aware of the limitations and problems of the NVD, they have a very positive attitude towards the NVD, as the database is helpful in obtaining information on vulnerabilities and provide guidance on vulnerability management in general. However, there is still room for improvement. For example, users would like to see a better check in quality of NVD records 
and CNAs, as well as better integration of NVD information into their own systems. The NVD is already working on implementing these suggestions, but is constrained by limited resources.
One negatively received aspect is the US government's control over the NVD and the limited input from other parties, as the government can influence the CVE List and the NVD by controlling their budget.
Users would like to see a more open, crowd-sourced platform where the quality of NVD records could be improved in a decentralized fashion. The NVD is working on this with introducing an NVD Consortium \citep{nvd-consortium}, 
which organizations can join if they meet certain requirements. With this, external parties would be able to participate in the NVD, which might lead to the necessary improvement of the database. 

Overall, the participants were satisfied with the database and appreciate it. As a result, the NVD will most likely continue to be heavily involved in day-to-day work, even with the problems we identified. The planned improvements to the NVD sound promising and are the first step towards long-term improvement of the valuable database.

\subsection{Limitations}
One limitation of the paper is the unbalanced demographic composition of the participants, as almost all participants are male. However, it is unknown what the demographics of regular NVD users is. Only people who use the NVD on a regular basis were surveyed. 
People who strictly decided to not use the NVD 
were not questioned. With around 71 participants, it is not possible to make a representative statement with respect to the population of NVD users. 
Nevertheless, we 
provide a general overview of the problems encountered by NVD users and their attitude towards the NVD.
\section{Conclusion}
\label{sec:conclusion}

In this paper, we conducted user studies to investigate how the NVD is utilized by users in everyday life, what problems users encounter and what their personal attitude towards the NVD is. For the qualitative preliminary study, a total of seven participants were interviewed. This was followed by a quantitative online survey with 71 participants. In order to investigate the root of the problems identified in the studies, we discussed the results with two senior NVD members.

The results show that the NVD is consulted almost daily and often serves as the main source to obtain a reference for vulnerability assessments or an overview of related patches or updates. The information then mainly flows into internal company tools and is used for further vulnerability processing. The information also feeds into users' own security advisories and is passed on to customers. Overall, users view the NVD very positively and praise its well-structured and standardized architecture. Nevertheless, there are also problems that users encounter from time to time. For example, users complain about incomplete or incorrect NVD records. The information, such as the description of the vulnerability and the corresponding CVSS score, often do not match.
In addition, there are sometimes very long delays from the discovery of the vulnerability until it is finally published. Due to these problems, the NVD is typically only used as a starting point to gather additional information about a vulnerability.

NVD members are aware of these problems. 
Many issues arise because the quality of information on
vulnerabilities supplied by vendors varies greatly.
This results in gaps in CVE List entries.
The NVD takes and enriches the entries by e.g. linking advisories, which can be very time-consuming and might also lead to delays.
However, other information, such as incomplete vulnerability descriptions are adopted directly.
One solution could be a standardized language for vulnerability descriptions and mandatory entries as most of them are optional.
The NVD is working on these problems, although restricted by its limited
resources.

The results show that most of the problems are not caused directly by the NVD, as it only passes on the information as an aggregator. 
Some problems, such as incomplete entries or inadequate descriptions arise in higher-level structures such as the CVE List, over which the NVD has no influence. 
A future study could interview members of the CVE Program about the root causes of the found problems. Another study could also investigate why there are variations in the quality of different CNAs or quantitatively research which CNAs are prone to making errors.

Besides the NVD, there are other databases that provide information on vulnerabilities. In a future study, these databases could be analyzed for their reliability and compared with the NVD.
Another future study could investigate which efforts have been applied by the NVD to overcome the known issues found in this study. 
Other aspects, such as recent changes to the CVE and NVD system, for example the automated addition of CVE records, could also be investigated.
Many of the problems could be solved with standardized formats and procedures, such as a uniform language for vulnerability descriptions. Future work could take on this task and develop a new approach or test the usability of existing projects and formats. 
In conclusion, a new standard only makes sense if it can solve existing problems and maintain the beneficial aspects of the NVD.

\paragraph{Acknowledgements}
We want to thank Thomas Schreck and Art Manion for establishing contact with Tanya Brewer and Christopher Turner for us. We would also like to thank Tanya Brewer and Christopher Turner, for being so open with use while discussing our results.
We also want to thank the testers and participants of the survey. 
We would like to thank the anonymous reviewers who improved the paper with their valuable feedback. This work was partially funded by the German
Federal Ministry of Education and Research under grant 16KIS1271K and by Deutsche Forschungsgemeinschaft (DFG, German Research Foundation) as part
of the Research and Training Group 2475 ``Cybercrime and Forensic Computing'' (grant number 393541319/GRK2475/2-2024). 

\paragraph{Author Contribution Statements}
\textbf{Julia Wunder:} Conceptualization, Methodology, Formal analysis, Investigation, Data Curation, Writing - Original Draft (Chapters 2.2, 2.3, 4, 6, 7), Writing - Review \& Editing, Visualization, Supervision, Project administration
\textbf{Alan Corona:} Conceptualization, Methodology, Formal analysis, Investigation, Data Curation, Writing - Original Draft (Chapters 2.1, 3, 5), Writing - Review \& Editing
\textbf{Andreas Hammer:} Formal analysis, Writing - Original Draft (Chapter 1), Writing - Review \& Editing
\textbf{Zinaida Benenson:} Writing - Review \& Editing, Funding acquisition
	
\bibliographystyle{ACM-Reference-Format}
\bibliography{nvd-paper}
	
\end{document}